\documentclass[prb,twocolumn]{revtex4}
\usepackage[dvips]{graphicx}
\usepackage{latexsym}
\usepackage{bm}
\usepackage{wasysym}

\def\etal{~\textit{et~al.}} 
\def\ra{\rangle} 
\def\la{\langle} 
\def\const{{\rm const}}
\def\Hc{{\rm h.c.}}

\def\mod{\rm{mod}}

\begin{document}

\title{Bosonic model with $Z_3$ fractionalization}
\author{O. I. Motrunich}
\affiliation{Massachusetts Institute of Technology,
77 Massachusetts Ave.,  Cambridge, MA 02139}

\date{October 30, 2002}

\begin{abstract}
Bosonic model with unfrustrated hopping and short-range repulsive
interaction is constructed that realizes $Z_3$ fractionalized 
insulator phase in two dimensions and in zero magnetic field.  
Such phase is characterized as having gapped charged excitations 
that carry fractional electrical charge $1/3$ and also gapped 
$Z_3$ vortices above the topologically ordered ground state.
\end{abstract}

\maketitle

\section{Introduction}
A flurry of recent theoretical activity has produced specific
model system realizations of fractionalized phases in two 
dimensions.\cite{RSSpN, MoeSon, Iof, BalMPAFGir, frcmdl, bosfrc}
Essentially all of the fractionalized states constructed so far are 
$Z_2$ states.\cite{U1}
On a formal level, these realizations employ the following route
to $Z_2$ fractionalization:  Strong local correlations lead to
a $U(1)$ gauge theory as a low-energy description; this gauge theory
is then driven into a deconfined state by a condensation of objects 
carrying gauge charge $2$ that also appear in the low-energy 
description.  This formal structure has been brought out very 
directly in Refs.~\onlinecite{SV, frcmdl, bosfrc}.
On a more physical level, the fractionalized insulator is produced 
departing from a superconducting state by a condensation of double 
vortices.\cite{NLII, z2long}
The main body of work concentrated on the $Z_2$ states 
since these are expected to be the simplest to realize.
However, it is clear that more complicated fractionalized 
states are also possible.
For example, it is conceivable that in some system the superconducting
state is quantum-disordered by a condensation of triple vortices; 
the resulting insulator is then a $Z_3$ fractionalized state.

In this paper, we indicate how a $Z_3$ fractionalized state
can be engineered in a relatively simple bosonic model with
unfrustrated nearest-neighbor hopping and short-range two-body
repulsive interaction.  Much of the construction parallels closely 
the $Z_2$ examples of Refs.~\onlinecite{frcmdl, bosfrc, Iof}:
The low-energy Hilbert space is selected---by stipulating 
particular charge interactions---in a manner that naturally 
admits splitting boson charge into three pieces; 
this Hilbert space is protected by a large charge gap.
The effective description of the fractionalized state has gapped 
chargons carrying electrical charge $1/3$ and coupled to 
some special $Z_3$ gauge theory which we analyze in detail.
Our main message here is that one does not need very contrived 
systems to obtain more complicated fractionalization patterns.

\section{$Z_3$ via charge frustration}
\begin{figure}
\centerline{\includegraphics[width=2.6in]{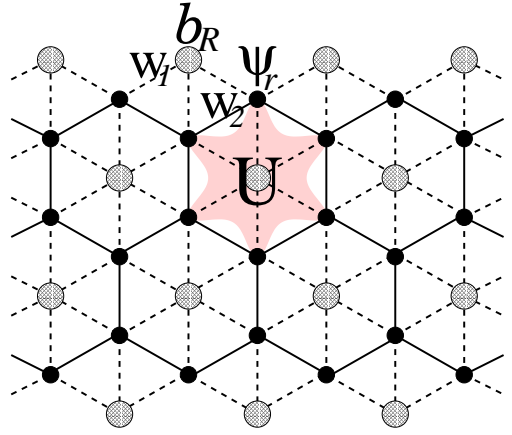}}
\vskip -2mm
\caption{Josephson junction array formed by penetrating 
hexagonal ($r$) and triangular ($R$) lattices, modeled by the 
Hamiltonian Eq.~(\ref{Hboson}).
The shaded area indicates schematically hexagon charging 
energy $U N_R^2$.
}
\label{lattice}
\end{figure}

The model is defined on the lattice shown in Fig.~\ref{lattice},
which we can think of as a hexagonal lattice with additional
sites placed at the hexagon centers.  
We have $\psi_r^\dagger = e^{i\phi_r}$ bosons residing on 
the hexagonal lattice (these sites are always labelled 
lower-case $r$), and $b_R^\dagger = e^{i\theta_R}$ bosons 
residing at the hexagon centers (upper-case $R$).
Bosons can hop between the neighboring sites as indicated by
the links on the figure; the hopping amplitudes are $w_1$
for $\la r, R \ra$ links and $w_2$ for $\la r,r' \ra$ links.
We also stipulate strong repulsive interactions that favor
charge neutrality of the hexagons, in addition to the on-site 
repulsion that favors charge neutrality of the individual sites.
The complete {\it quantum rotor} Hamiltonian is
\begin{eqnarray}
\label{Hboson}
H & = &
-w_1 \sum_{R, r \in R} (b_R^\dagger \psi_r + \Hc)
-w_2 \sum_{\la rr' \ra} (\psi_r^\dagger \psi_{r'} + \Hc) 
\nonumber \\
&& + u_b \sum_R (n_R^b)^2 + u_\psi \sum_r (n_r^\psi)^2 
   + U \sum_R N_R^2 ~.
\end{eqnarray}
Here, $\{n_R^b, \theta_R\}$ are conjugate number-phase variables 
[e.g., in the phase representation
$n_R^b \!\equiv\! -i\partial/\partial{\theta_R}$, 
$b_R^\dagger \psi_r + \Hc \!\equiv\! 2\cos(\theta_R-\phi_r)$],
and similarly for $\{n_r^\psi, \phi_r\}$.
The number-phase variables are particularly appropriate if we think of 
the model as describing an array of Josephson-coupled superconducting 
islands.
Both $b_R$ and $\psi_r$ bosons carry electrical charge $q_b$.

In the above Hamiltonian, $N_R$ is the boson number associated
with each hexagon:
\begin{equation}
\label{NR}
N_R = 3 n_R^b + \sum_{r \in R} n_r^\psi ~.
\end{equation}
Thus, the total boson number in the system\cite{boundaries} is
\begin{equation}
\label{Ntot}
N_{\rm tot} = \frac{1}{3} \sum_R N_R.
\end{equation}

General analysis of the possible phases proceeds as in 
Ref.~\onlinecite{bosfrc}.  
Here, we focus on the insulating states that are obtained 
for large $U \gg w_1, w_2, u_b, u_\psi$.  
If the $w_1, w_2, u_b, u_\psi$ terms are all zero, there is a 
degenerate manifold of ground states specified by the requirement 
$N_R=0$ for each $R$.\cite{numbers}
The ground state sector is separated by a large charge gap $U$ from 
the nearest sectors.  Including the $w_1, w_2, u_b, u_\psi$ terms 
lifts this degeneracy in each sector, which is best described by 
deriving the corresponding effective Hamiltonian in each sector.

The effective Hamiltonian in the ground state sector ($N_R=0$) is,
to third order in the perturbing terms,
\begin{eqnarray}
\label{Heff}
H_{\rm eff}^{(0)} = H_{u_b, u_\psi}
-J_{\rm c} \sum_r
   \left[ (\psi_r^\dagger)^3 b_{R_1} b_{R_2} b_{R_3} + \Hc \right]
\nonumber \\
-K_{\rm ring} \sum_{\hexagon}
   \left( \psi_1^\dagger \psi_2 \psi_3^\dagger \psi_4 
          \psi_5^\dagger \psi_6 + \Hc \right) ~.
\end{eqnarray}
Here, $H_{u_b,u_\psi}$ stands for the on-site repulsion terms
as in Eq.~(\ref{Hboson});
$R_1, R_2, R_3$ label the three hexagon centers adjacent to $r$;
$J_{\rm c}=w_1^3/(6U^2)$ and $K_{\rm ring}=3 w_2^3/U^2$.

This is our main step in obtaining $Z_3$ fractionalization.  
The above Hamiltonian looks similar to a compact $U(1)$ gauge theory 
coupled to a charge $3$ scalar field.  Thus, if we think of the 
$\psi_r$ as some gauge fields, then it is very suggestive to think of 
the $b_R$ as carrying gauge charge $3$ [see also Eq.~(\ref{NR})].
Standard Fradkin-Shenker analysis\cite{FraShe} then suggests that
by condensing the $b_R$ field, which can be arranged by making
$J_{\rm c}, K_{\rm ring}$ large, we can deconfine objects carrying 
gauge charge $1$.  By virtue of Eq.~(\ref{Ntot}), such objects carry 
fractional electrical charge $q_b/3$, and we obtain a $Z_3$ 
fractionalized insulator.

The reader who finds the above statements believable may now
declare victory in achieving $Z_3$ fractionalization.
However, if we want to describe the deconfined phase(s) in 
any detail, we need to study the above Hamiltonian directly 
since it is not related in any simple manner to the conventional 
gauge theory.  This is our focus in the remainder of the paper.
Of course, we will confirm the deconfinement, but we will
also find that the deconfined state that obtains for large
$K_{\rm ring}$ on all hexagons is in fact a $Z_3 \times Z_3$
state (see below for details).

Proceeding with this analysis, consider the regime of large 
$J_{\rm c} \to \infty$ and small $u_b \to 0$.
It is convenient to perform the following change of variables.
Define the operators $b_{cR}^\dagger = e^{i\theta_{cR}}$ and 
$\tilde\psi_r^\dagger = e^{i\tilde\phi_r}$:
\begin{equation}
b_{cR}^\dagger = s_R e^{i\theta_R/3}, \quad
\tilde\psi_r^\dagger = \psi_r^\dagger b_{cR_1} b_{cR_2} b_{cR_3} ~. 
\end{equation}
Here, $s_R = 1, e^{i 2\pi/3}$, or $e^{i 4\pi/3}$, so that
$\theta_{cR} \in [0, 2\pi)$.
One can readily verify that $N_R$ is conjugate to $\theta_{cR}$,
while $n_r^\psi$ is conjugate to $\tilde\phi_r$.
$b_{cR}^\dagger$ carries electrical charge $q_b/3$ and can be 
thought of as a chargon field, while $\tilde\psi_r^\dagger$ is 
charge neutral.
These new variables are indeed natural in the description of the
deconfined phase, but to recover the physical Hilbert space,
we need to impose the constraint
\begin{eqnarray}
\exp\left[i \frac{2\pi}{3} ( N_R-\sum_{r \in R} n_r^\psi ) \right] 
= 1 ~.
\end{eqnarray}
In the new variables, the Hamiltonian becomes
\begin{eqnarray}
H_{\rm eff}^{(0)} =  u_\psi \sum_r (n_r^\psi)^2 
-J_{\rm c} \sum_r
   \left[ (\tilde\psi_r^\dagger)^3 + \Hc \right]
\nonumber \\
-K_{\rm ring} \sum_{\hexagon}
  \left(\tilde\psi_1^\dagger \tilde\psi_2 \tilde\psi_3^\dagger 
        \tilde\psi_4 \tilde\psi_5^\dagger \tilde\psi_6 + \Hc \right) ~.
\end{eqnarray}
Note that we are left with gauge fields $\tilde\psi_r$ only, 
since we are working in the ``uncharged'' ground state sector

The term $J_{\rm c}$ acts as a $Z_3$ anisotropy on the 
$\tilde\phi_r$ field.
In the limit $J_{\rm c} \to \infty$, $e^{i\tilde\phi_r}$ becomes a 
$Z_3$ field: $\tilde\phi_r = 0, 2\pi/3$, or $4\pi/3$.
The operator $P_r^{+} \equiv e^{- i (2\pi/3) n_r^\psi}$ shifts the 
states of the quantum $Z_3$ clock by $+1$,\cite{ZN} whereas the 
constraints specifying the uncharged sector $N_R=0$ become
\begin{eqnarray}
\label{Puncharged}
\prod_{r\in R} P_r^{+} = 1 ~.
\end{eqnarray}
The $u_\psi (n_r^\psi)^2$ term causes tunneling between the different
states of the quantum clock, and this can be described by an effective
``transverse field'' $-h (P_r^+  + P_r)$ acting on the clock.

\begin{figure}
\centerline{\includegraphics[width=2.6in]{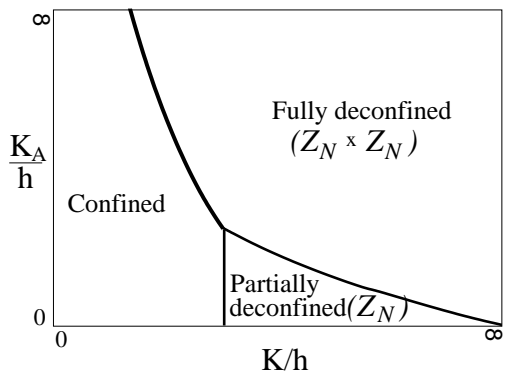}}
\vskip -2mm
\caption{Generic phase diagram of the $Z_N$ ring exchange on hexagonal 
lattice, Eq.~(\ref{Hhex}).  To explore the different deconfined
states, we allow two different ring exchange couplings, 
$K_{\rm ring}=K_A$ for {\tt A}-type hexagons and 
$K_{\rm ring}=K$ for {\tt B}- and {\tt C}-type hexagons
(see Fig.~\ref{hexlattice}). 
}
\label{phased}
\end{figure}

The effective Hamiltonian now becomes $Z_3$ ``ring-exchange'' 
Hamiltonian on the hexagonal lattice of $r$ sites
\begin{eqnarray}
\label{Hhex}
-K_{\rm ring} \sum_{\hexagon}
   \left( \tilde\psi_1^\dagger \tilde\psi_2 \tilde\psi_3^\dagger 
          \tilde\psi_4 \tilde\psi_5^\dagger \tilde\psi_6 + \Hc \right)
-h \sum_r (P_r^{+} + P_r^{-}) ~.
\end{eqnarray}
The Hamiltonian together with the constraints Eq.~(\ref{Puncharged})
can be viewed as some special $Z_3$ gauge theory and is analyzed
below and in further detail in 
Appendices~\ref{app:ringexch}~and~\ref{app:dual}.  
We find that generically this theory can have {\it two} deconfined 
phases (in addition to the confined phase) with the phase diagram 
shown in Fig.~\ref{phased}.
Here we only describe the deconfined phase that obtains when all 
ring exchange couplings are large, $K_{\rm ring} \gg h$.  
As explained below, this phase is a $Z_3 \times Z_3$ deconfined phase.

From here on, our focus is on the above $Z_3$ ring exchange 
Hamiltonian.  We drop all tildes on $Z_3$ fields $\psi_r$ and 
superscripts on $n_r$ (which are now integers modulo $3$).
Also, we consider a $Z_N$ generalization\cite{ZN} of the above 
Hamiltonian and carry out the analysis in the general case.
This is done for clarity of notation.  

For $K_{\rm ring} \gg h$, a good caricature of the bulk ground 
state is given by the wavefunction
\begin{eqnarray}
\label{GS}
|GS\ra = \sum_{\{n_r\}} \! ^\prime \;
|\{n_r\}\ra ~,
\end{eqnarray}
where the primed sum is over all configurations $\{n_r\}$ 
that satisfy the constraints Eq.~(\ref{Puncharged}),
i.e., $\sum_{r\in R} n_r = 0$.

\begin{figure}
\centerline{\includegraphics[width=\columnwidth]{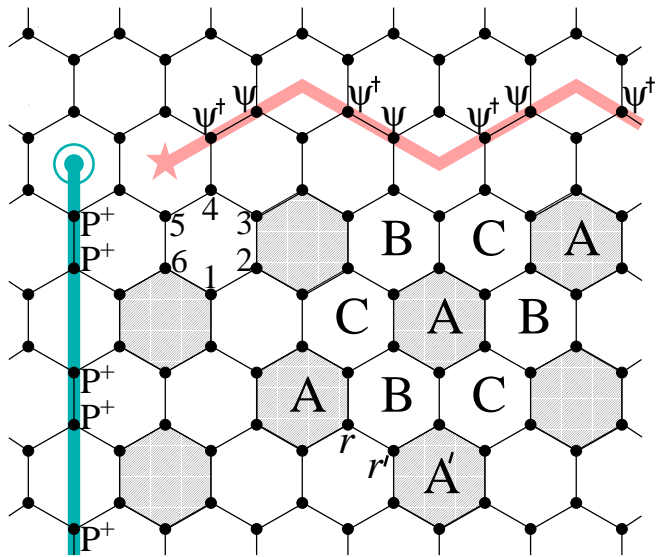}}
\vskip -2mm
\caption{Hexagonal lattice on which the $Z_N$ ring exchange model
Eq.~(\ref{Hhex}) is defined.  
$N_R \equiv \sum_{r\in R} n_r$ measures $Z_N$ ``number'' on each 
hexagon; we can increase $N_R$ on a given hexagon ($\star$) by one by 
applying a ``string'' operator 
$\psi^\dagger \psi \psi^\dagger \psi \dots$ along the indicated path.
$\Phi_R \equiv \phi_1 - \phi_2 + \phi_3 - \phi_4 + \phi_5 - \phi_6$
measures $Z_N$ ``flux'' through a given hexagon $(123456)$;
to fix the sign convention we always take $1$ to be the lowermost 
hexagon site.  We can increase the flux through a given hexagon
($\bullet$) by one unit by applying a string
$P^{+} P^{+} P^{+} P^{+} \dots$ along the indicated
path.  The lower-right corner of the figure shows the three sublattice 
structure of the lattice of honeycombs.
}
\label{hexlattice}
\end{figure}

Let us define $Z_N$ flux through a given hexagon $R$
\begin{eqnarray}
\label{PhiR}
\Phi_R = \phi_1 - \phi_2 + \phi_3 - \phi_4 + \phi_5 - \phi_6 ~,
\end{eqnarray}
with the sign convention as in Fig.~\ref{hexlattice}.
The ground state has zero flux through each hexagon.
Excitations above this ground state are $Z_N$ vortices.
For example, we can add a unit of flux through a given hexagon by 
applying a ``string'' operator as indicated in Fig.~\ref{hexlattice}.
The gap for a vortex carrying one unit of flux is
$2 K_{\rm ring} [1-\cos(2\pi/N)]$.

Observe now (Fig.~\ref{hexlattice}) that the hexagon centers $R$ 
form a triangular lattice, which consists of three sublattices 
{\tt A}, {\tt B}, and {\tt C}.  Observe also that the 
flux-adding string operator ``steps'' only through the same 
sublattice hexagons.  We are thus led to the possibility of a 
topological distinction between vortices on the different 
sublattices, in addition to the usual distinction between two 
vortices carrying different flux.
Indeed, one can see that the topologically distinct situations
can be characterized by saying that we have two species of
$Z_N$ vortices, say {\tt A} and {\tt B} vortices.  
Alternatively, if we want to preserve the symmetry among the three 
sublattices, we can say that there are three types of $Z_N$ 
vortices---{\tt A}, {\tt B}, and {\tt C} vortices---but these are not 
independent and instead satisfy ``fusion rules'' such as
\begin{eqnarray}
\label{fusion:fluxes}
\Big(\Phi_A \!=\! +1 \Big) \times \Big(\Phi_B \!=\! +1 \Big) 
\;\;\sim\;\; \Big(\Phi_C \!=\! -1 \Big) ~.
\end{eqnarray}
This means that a nearby pair of $+1$ {\tt A} and {\tt B}
vortices is indistinguishable from a $-1$ {\tt C} vortex
(note also that the ``states'' on the left and on the right can be
connected by local $h$ terms in the Hamiltonian).

Consider now introducing $Z_N$ charges in the above gauge theory
Eq.~(\ref{Hhex}), e.g., consider placing a pair of opposite $\pm 1$
charges on two hexagons $R_1$ and $R_2$:
$\sum_{r\in R} n_r = \delta_{RR_1} - \delta_{RR_2}$.
This is appropriate when studying the charged sectors of
the microscopic Hamiltonian Eq.~(\ref{Hboson}) since the formal
gauge structure represents the crucial coupling of chargons with 
the above $Z_3$ degrees of freedom.
A charge can be added to a hexagon by applying a string operator 
as indicated in Fig.~\ref{hexlattice}.
From several perspectives, one can see that all such charges
are deconfined in the $K_{\rm ring} \gg h$ phase:  
Thus, in Appendix~\ref{app:ringexch} we approach 
this ``fully deconfined'' phase starting from a 
``partially deconfined'' phase, in which charges are deconfined
on one sublattice only.  Also, this fully deconfined
phase corresponds to the fully disordered phase in the dual 
global $Z_N$ spin model of Appendix~\ref{app:dual}.

Similarly to vortices, we need to distinguish the charges on 
different sublattices.  Again, as far as the gauge structure
is concerned, we have fusion rules such as 
\begin{eqnarray}
\label{fusion:charges}
\Big(N_A \!=\! +1 \Big) \times \Big(N_B \!=\! +1 \Big) 
\;\;\sim\;\; \Big(N_C \!=\! -1 \Big) ~.
\end{eqnarray}

Statistical interactions between the different particles
are readily identified by studying the commutation properties 
of the corresponding strings.  These are summarized in 
Table~\ref{tab:charges} by specifying ``gauge charges'' of the
different $N_R \!=\! +1$ excitations with respect to the 
{\tt A}, {\tt B}, and {\tt C} fluxes.
Thus, $N_A \!=\! +1$ excitation carries gauge charges
$Q_A=0, Q_B=-1, Q_C=+1$, i.e., it does not ``see'' {\tt A} vortices, 
but when transported around a {\tt B} or {\tt C} vortex of unit 
strength, the wavefunction acquires an additional phase 
$e^{-i 2\pi/N}$ or $e^{i 2\pi/N}$ correspondingly.

\begin{table}
\vskip -2mm
\caption{\label{tab:charges} Gauge charges of the $N_R = +1$
excitations with respect to the {\tt A}, {\tt B}, and {\tt C}
$Z_N$ fluxes as defined by Eq.~(\ref{PhiR}).  Note that these
are consistent with the fusion rules 
Eqs.~(\ref{fusion:fluxes}) and (\ref{fusion:charges}).
}
\begin{ruledtabular}
\begin{tabular}{lrrr}
 & $Q_A$  & $Q_B$ & $Q_C$\\
\hline
$\Big(N_A \!=\! +1 \Big)$ & $ 0$  & $-1$  &  $+1$ \\
$\Big(N_B \!=\! +1 \Big)$ & $+1$  & $ 0$  &  $-1$ \\
$\Big(N_C \!=\! +1 \Big)$ & $-1$  & $+1$  &  $ 0$
\end{tabular}
\end{ruledtabular}
\vskip -2mm
\end{table}

This completes the particle description of the fully
deconfined phase.  The minimal description would be to say
that we have {\tt A} hexagon charges that see {\tt B} hexagon 
$Z_N$ vortices and {\tt B} hexagon charges that see {\tt A} hexagon 
vortices.  This is essentially the claimed $Z_N \times Z_N$ structure.
Thus, we expect $N^2$-fold ground state degeneracy if the system
is put on a cylinder, as can be verified by constructing the 
corresponding ground states starting from the state Eq.~(\ref{GS}) 
and threading vortices through the hole of the cylinder.

It should be emphasized here that the above discussion
assumed that the three-sublattice structure is respected by 
the boundary conditions.  While it is clear that the bulk
properties do not depend on this, there is an additional 
quirk when we consider topological degeneracy in a geometry that 
does not respect the three-sublattice structure.
This is legitimate when all ring exchange couplings are equal.
Consider, e.g., a cylindrical geometry with the circumference
along the horizontal direction of 
Figs.~\ref{lattice}~and~\ref{hexlattice}.
When an {\tt A}-type particle is transported around the periodic 
direction, it does not return to its initial position, but rather 
becomes a {\tt B}- or {\tt C}-type particle.  It takes three turns 
for the particle to return to the original position.  
From table~\ref{tab:charges}, such $N_R=+1$ particle will not 
register any flux in this process.  
A detailed analysis shows that for $N \;\mod\; 3 \neq 0$ the 
ground state of the system in this geometry is nondegenerate.
On the other hand, for $N \;\mod\; 3 = 0$ the ground state is 
found to be three-fold degenerate, since in this case there is a 
composite object that returns to its initial state when transported 
once around the cylinder and that senses some flux through the hole 
of the cylinder in the process.

Returning to our microscopic bosonic model, the 
``particle description'' of the $Z_3 \times Z_3$ phase is as follows: 
We have two species of $Z_3$ vortices (with gap $\sim K_{\rm ring}$) 
and we have charged particles (with charge gap $\sim U$) that can be 
classified as carrying two distinct $Z_3$ gauge charges, in addition
to their fractional electrical charge.
Finally, note that the $Z_3 \times Z_3$ state is associated with 
the additional symmetries in the hexagonal lattice ring exchange
Hamiltonian but is protected by the same charge gap projection, 
since any move within the uncharged state sector is necessarily a 
combination of hexagon ring exchanges.

\section{Conclusions}
We showed that it is possible to produce more complicated 
fractionalization patterns such as $Z_3$ fractionalization
in relatively simple bosonic models.
While the resulting fractionalized state turned out to be even 
more complicated than initially intended, the microscopic model 
was not too contrived.  It is hoped that this work will 
encourage further searches for other exotic states.  
For example, can a non-Abelian fractionalized state be produced in a 
condensed matter system with a global symmetry only, short-range 
interactions, and in zero magnetic field?

\acknowledgments
The author particularly thanks T. Senthil for many stimulating 
discussions and persistent encouragement that this toy project
be written up, and also for useful comments on the manuscript.
This work was supported by the MRSEC program of 
the National Science Foundation under grants DMR-9808941
and DMR-0213282.

\appendix
\section{Partially deconfined ($Z_N$) phase}
\label{app:ringexch}
To better appreciate the character of the deconfinement
in the special gauge theory Eq.~(\ref{Hhex}), we allow
different ring exchange couplings for different hexagons
and consider particular parameter space with two such couplings: 
$K_{\rm ring}=K_A$ for the {\tt A} hexagons and 
$K_{\rm ring}=K_B=K_C\equiv K$ for the {\tt B} and {\tt C} hexagons.  
This is indicated schematically in Fig.~\ref{hexlattice} where the 
{\tt A} hexagons are shaded.  Note that by allowing the two couplings
we implicitly assume that the boundary conditions on the lattice 
respect the three-sublattice structure; this is done throughout.

We argue below that the ring exchange Hamiltonian has the 
phase diagram shown in Fig.~\ref{phased} with three phases:  
For $h \gg K, K_A$ the system is in a fully confined phase.
For $K \gg h \gg \sqrt{K_A K}$ the system is in a partially deconfined
($Z_N$) phase.  In this phase, the charges on the {\tt A} hexagons are 
deconfined, while the charges on the {\tt B} and {\tt C} hexagons are 
confined.
Finally, for $K, K_A \gg h$ the system is in a fully deconfined 
($Z_N \times Z_N$) phase with all charges deconfined.
The phase diagram of Fig.~\ref{phased} is also supported by
the analysis of the dual global $Z_N$ spin model summarized in 
Appendix~\ref{app:dual}.

In what follows, we give a detailed description of the partially 
deconfined phase.
As a representative of this phase, consider the Hamiltonian with 
$K_A=0$, i.e., with ring exchanges around the
{\tt B} and {\tt C} hexagons only (see Fig.\ref{hexlattice}).  
In this case, there are additional conserved quantities:
\begin{eqnarray}
{\hat{\cal L}}_{AA'} \equiv n_r + n_{r'} = \const \quad \quad 
\text{(model with $K_A = 0$)}
\end{eqnarray}
for each hexagonal lattice link $\la rr' \ra$ between two 
{\tt A} hexagons $A$ and $A'$ (see Fig.~\ref{hexlattice}).  
This facilitates the analysis, since we can consider separately each 
subsector specified by the corresponding eigenvalues 
$\{{\cal L}_{AA'}\}$.  Note that the allowed 
$\{{\cal L}_{AA'}\}$ are very much constrained by the 
constraints Eq.~(\ref{Puncharged}) on the $n_r$ themselves; 
however, we will not use the details of these explicitly.

First of all, observe that the {\tt A} hexagons in turn form a 
triangular lattice, while the links $\la rr' \ra$ between such 
hexagons can also be viewed as the links of this ``{\tt A}-lattice'',
$\la AA' \ra \equiv \la rr' \ra$.
In a given subsector with fixed $\{{\cal L}_{AA'}\}$, there remains 
one $Z_N$ degree of freedom for each such link.  It is convenient to 
work in the number basis and label these remaining link degrees of 
freedom by
\begin{eqnarray}
\label{NAA}
{\cal N}_{A \to A'} \equiv n_r-n_r^{(0)} 
= -(n_{r'}-n_{r'}^{(0)}) \equiv -{\cal N}_{A' \to A},
\end{eqnarray}
where $\{ n_r^{(0)} \}$ is one particular instance:
${\cal L}_{AA'} = n_r^{(0)} + n_{r'}^{(0)}$ (and our
convention is that $r \in A$ and $r' \in A'$---see 
Fig.~\ref{hexlattice}).
Thus, ${\cal N}_{AA'} \equiv {\cal N}_{A \to A'}$ are oriented
fields on the links of the {\tt A}-lattice.  
The subsector is now completely specified by the conditions
\begin{eqnarray}
\label{NAAneutral}
\sum_{A' \in A} {\cal N}_{AA'} = 0 ~,
\end{eqnarray}
which are the neutrality constraints Eq.~(\ref{Puncharged})
for the {\tt A} hexagons.

The action of the Hamiltonian Eq.~(\ref{Hhex}) in this subsector 
is readily described in terms of the new variables.  
Thus, the transverse field ($h$) terms are diagonal in the 
new number variables, while the {\tt B} and {\tt C} hexagon ring 
exchanges simultaneously raise (or lower) the three oriented number 
fields circulating around the corresponding {\tt A}-lattice 
triangular plackets.  
Writing the raising operator for a given link number variable
${\cal N}_{AA'}$ as $e^{i\Xi_{AA'}}$, the resulting 
Hamiltonian is
\begin{eqnarray}
\label{HtriGT}
\hat H[{\cal L}] = 
-K\sum_{\triangle} 
     \left(e^{i\Xi_{AA'}} e^{i\Xi_{A'A''}} e^{i\Xi_{A''A}} + \Hc\right)
\\ \nonumber
-\sum_{\la AA' \ra} 
     \left(\Gamma_{AA'} e^{-i (2\pi/N) {\cal N}_{AA'}} + \Hc \right) ~,
\end{eqnarray}
where
\begin{eqnarray}
\Gamma_{AA'} = h e^{-i (2\pi/N) n_r^{(0)}} 
                 (1 + e^{i (2\pi/N) {\cal L}_{AA'}}) ~.
\end{eqnarray}
Together with the constraints Eq.~(\ref{NAAneutral}), 
this is precisely the conventional $Z_N$ lattice gauge theory 
defined on the triangular {\tt A}-lattice but with link-dependent 
$\Gamma_{AA'}$ specific for the particular subsector 
$\{{\cal L}_{AA'}\}$.
We can now use the conventional wisdom to characterize each
such subsector and in turn the full hexagonal ring
exchange Hamiltonian with $K_A=0$.

When $h\equiv 0$, all the different subsectors are degenerate.
The lowest energy state in each such subsector has the energy of 
$-2K$ per triangle and is an equal weight superposition of all 
possible configurations of ${\cal N}_{AA'}$ that satisfy the 
constraints Eq.~(\ref{NAAneutral}).
Nonzero $h$ eliminates this degeneracy and selects one particular 
subsector, namely with all ${\cal L}_{AA'} = 0$, as containing the 
true ground state of the full Hamiltonian with $K_A=0$.  
Indeed, treating $\Gamma_{AA'}$ perturbatively, the lowest energy 
in a given subsector is
\begin{eqnarray}
E_{GS}[{\cal L}] \approx -\sum_\triangle 2K 
-\sum_{\la AA' \ra} \frac{ |\Gamma_{AA'}|^2 }{ 2 K[1-\cos(2\pi/N)] } ~,
\end{eqnarray}
where for simplicity we assumed that the system has no boundaries.
It is now clear that for small nonzero $h$ the ground state of the 
full ring exchange Hamiltonian with $K_A=0$ is in the subsector with 
all ${\cal L}_{AA'}=0$.  The subsectors that are closest in energy 
have the smallest number of nonzero
${\cal L}_{AA'}$ and can be characterized as having alternating
${\cal L}_{A_0 A'}=+1$ and ${\cal L}_{A_0 A'}=-1$ values on the six
links to a given hexagon $A_0$ (this subsector is obtained from the 
ground state subsector by applying the hexagon ring exchange 
around the hexagon $A_0$).  The energy gap to these 
subsectors is $6 h^2/K$.  
We see that we have a peculiar situation where a nonzero 
transverse field $h$ is needed to stabilize this $Z_N$ deconfined 
ground state; this is because we are competing here against 
the $Z_N \times Z_N$ deconfined state that is obtained for 
large $K_A, K$.

We are all set to discuss confinement of charges in the model with 
$K_A=0$.  The above analysis was carried out in the uncharged sector 
but is readily extended to the charged sectors.  
First, consider placing a pair of opposite charges on two 
{\tt B} hexagons:
$\sum_{r\in B_1} n_r = +1$ and $\sum_{r\in B_2} n_r = -1$.
Proceeding exactly as before, one is led to consider different
subsectors (of this charged sector) specified by $\{{\cal L}_{AA'}\}$.
In each such subsector, the Hamiltonian has precisely the form 
Eq.~(\ref{HtriGT}) with the number variables satisfying precisely
the constraints Eq.~(\ref{NAAneutral}).  
All information about the two charges is encoded in the allowed 
configurations $\{{\cal L}_{AA'}\}$, and one can clearly see that 
${\cal L}_{AA'} \neq 0$ at least on a string of {\tt A}-lattice
bonds connecting $B_1$ and $B_2$.  From the earlier arguments,
the energy cost of introducing two such charges is then proportional
to the length of this string, i.e., such charges are confined
with the string tension $\sim h^2/K$.

Consider now placing a pair of opposite charges on two 
{\tt A} hexagons $A_1$ and $A_2$: 
$\sum_{r\in A_1} n_r = +1$ and $\sum_{r\in A_2} n_r = -1$.
The analysis of the subsectors $\{{\cal L}_{AA'}\}$ will be
somewhat different in this case.
For each such subsector in this charged sector there corresponds 
a subsector in the uncharged sector having exactly the same 
$\{{\cal L}_{AA'}\}$.  It is convenient to ``measure'' each 
charged subsector relative to the corresponding uncharged subsector.  
This is achieved by defining link variables ${\cal N}_{AA'}$ via
Eq.~(\ref{NAA}) using an uncharged instance 
$\{ n_r^{(0)} \}$ of $\{{\cal L}_{AA'}\}$ 
(i.e., $\sum_{r\in R} n_r^{(0)} = 0$ for each $R$ and 
${\cal L}_{AA'} = n_r^{(0)} + n_{r'}^{(0)}$ for each 
$\la AA' \ra$).  In each subsector, the Hamiltonian again
has the form Eq.~(\ref{HtriGT}) when written in these link variables,
which now satisfy new constraints 
$\sum_{A'\in A_1} {\cal N}_{A_1 A'} = +1$ 
and $\sum_{A'\in A_2} {\cal N}_{A_2 A'} = -1$.  
This corresponding precisely to introducing two charges in the 
corresponding {\tt A}-lattice gauge theory.
Clearly, for large enough $K \gg h$, these charges will
be deconfined.

We now have essentially complete description of the 
partially deconfined phase.  Thus, one can readily identify
the $Z_N$ vortex excitations of the {\tt A}-lattice gauge theory
with $Z_N$ vortices on the {\tt B} and {\tt C} hexagons.
These vortices will have usual statistical interactions
with the deconfined charges on the {\tt A} hexagons.
Also, as should become clear by reviewing the above discussion,
we can essentially account for the different subsectors 
$\{{\cal L}_{AA'}\}$ by saying that there are additional particle 
excitations living on the {\tt A} hexagons obtained from the 
ground state by the action of the corresponding {\tt A} hexagon 
ring exchanges.  These new particles have a ``mass'' of $6h^2/K$ and 
have no statistical interaction with the other particles.

We can now consider what happens when we allow nonzero 
$K_A$.  As discussed above, the ring exchanges around the {\tt A}
hexagons introduce mixing between the different subsectors.  
However, as long as $K_A$ is much smaller than the corresponding
gap $\sim h^2/K$, the partially deconfined phase survives
and is characterized by the same particle description.

Once $K_A$ is sufficiently large, the system enters the 
fully deconfined phase described in the main text.

\section{Dual global $Z_N$ spin model}
\label{app:dual}
Here we summarize dual perspective on the hexagonal lattice 
ring exchange Hamiltonian Eq.~(\ref{Hhex}).
We work directly in the Hamiltonian language.
Simple counting shows that the dimensionality of the physical 
Hilbert space is consistent with having one $Z_N$ degree of freedom 
per hexagon.  Let us define
\begin{eqnarray}
T_R^{-} \equiv \psi_1^\dagger \psi_2 \psi_3^\dagger \psi_4 
          \psi_5^\dagger \psi_6 ~, 
\quad\quad T_R^{+} \equiv (T_R^{-})^\dagger ~,
\end{eqnarray}
where we use the same sign convention as in Fig.~\ref{hexlattice}.
Let us also define
\begin{eqnarray}
V_R^\dagger \equiv \prod_{\to R} P^{+} ~,
\end{eqnarray}
where the product is along the vertical path that reaches $R$ 
as in Fig.~\ref{hexlattice}.  Note that the path ``steps'' 
through the same sublattice hexagons.  [If we were to take some
other such path, we would need to replace some $P^{+}$ with $P^{-}$.
The total product is path-independent due to 
constraints~Eq.~(\ref{Puncharged}).]
$V_R^\dagger$ can be thought of as a vortex creation operator.

We now interpret $V_R^\dagger$ as a $Z_N$ spin variable.
It is easy to verify that $T_R^{+}$ is the corresponding
conjugate variable (i.e., raising operator\cite{ZN}):
\begin{eqnarray}
V_R^\dagger T_R^{+} = e^{i 2\pi/N} T_R^{+} V_R^\dagger ~.
\end{eqnarray}
Also, we can readily ``solve'' for $P_r^{+}$:
\begin{eqnarray}
P_r^{+} = V_{R_1}^\dagger V_{R_2}^\dagger V_{R_3}^\dagger ~.
\end{eqnarray}
The dual Hamiltonian is
\begin{eqnarray}
H = -K \sum_{R} (T_R^{+} + \Hc) 
    -h \sum_\triangle (V_{R_1}^\dagger V_{R_2}^\dagger V_{R_3}^\dagger
                       + \Hc) ~,
\end{eqnarray}
which is a global $Z_N$ spin model with three-spin interactions.
A little thought shows that the model has in fact a 
$Z_N \times Z_N$ global symmetry corresponding to independent
global rotations of the spins on two of the three sublattices.
Note also that the three-spin interaction around triangles 
promotes ordering of the spins on the same sublattice.  
This is because two neighboring sites $A$ and $A'$ on the same 
sublattice share a $BC$ side in the respective triangle interactions 
$\triangle ABC$ and $\triangle A'BC$.

The global model clearly has a fully disordered phase for $K \gg h$.  
In the original ring exchange Hamiltonian, this corresponds to all 
vortices being gapped, and we obtain the $Z_N \times Z_N$ fully 
deconfined phase.  Varying the {\tt A} hexagon ring exchange coupling
$K_A$ independently, for sufficiently small $K_A$ and large $K$
the system can clearly order on the {\tt A} sublattice 
(i.e., {\tt A} vortices condense),  but remain disordered on the 
{\tt B} and {\tt C} sublattices.
This is our partially deconfined $Z_N$ phase.


\begin{thebibliography}{10}
\bibitem{RSSpN}
N.~Read and S.~Sachdev, Phys. Rev. Lett. {\bf 66}, 1773 (1991);
X.-G.~Wen, Phys. Rev. B {\bf 44}, 2664 (1991).

\bibitem{MoeSon}
R.~Moessner and S.~L.~Sondhi, Phys. Rev. Lett. {\bf 86}, 1881 (2001).

\bibitem{Iof}
L.~B.~Ioffe\etal, Nature {\bf 415}, 503 (2002).

\bibitem{BalMPAFGir}
L.~Balents, M.~P.~A.~Fisher, and S.~M.~Girvin, 
Phys. Rev. B {\bf 65}, 224412 (2002).

\bibitem{frcmdl}
T.~Senthil and O.~I.~Motrunich, cond-mat/0201320.

\bibitem{bosfrc}
O.~I.~Motrunich and T.~Senthil, cond-mat/0205170.

\bibitem{U1}
One notable example of non-$Z_2$ fractionalized state is $U(1)$ 
Coulomb phase in three dimensions, Ref.~\onlinecite{bosfrc}. 
[See also
X.-G.~Wen, Phys. Rev. Lett. {\bf 88}, 011602 (2002), and
cond-mat/0210040].

\bibitem{SV}
S.~Sachdev and M.~Vojta, J. Phys. Soc. Jpn. {\bf 69}, 
Supp. B, 1 (2000).

\bibitem{NLII} 
L.~Balents, M.~P.~A.~Fisher, and C.~Nayak, 
Phys. Rev. B {\bf 60}, 1654 (1999); 
{\em ibid.} {\bf 61}, 6307 (2000).

\bibitem{z2long}
T.~Senthil and M.~P.~A.~Fisher,  Phys. Rev. B {\bf 62}, 7850 (2000).

\bibitem{numbers}
Recall that the operators $n_R^b$ and $n_{rr'}^{\psi}$ are defined as 
conjugates of the corresponding phase variables and have eigenvalues 
that can take all integer values including negative ones; 
thus, the constrained Hilbert space $N_R=0$ is indeed nontrivial.

\bibitem{boundaries}
To avoid any complications at boundaries when such are present, 
we require that each $r$ site has precisely three $R$ neighbors
$R_1, R_2, R_3$.  We can construct such an array, e.g., by starting 
from a triangular lattice of the $R$ sites, possibly with boundaries,
and placing the $r$ sites at the centers of the triangles.


\bibitem{FraShe}
E.~Fradkin and S.~H.~Shenker, Phys. Rev. D {\bf 19}, 3682 (1979).

\bibitem{ZN}
Each $Z_N$ variable is represented by
$\psi^\dagger = e^{i\phi} = e^{i 2\pi m/N}$, $m=0,1,\dots,N-1$.
$P^{+}$ is the raising operator on the phase $\phi$ and
is defined by the commutation relation
$ e^{i\hat\phi} P^{+} = e^{i 2\pi/N} P^{+} e^{i\hat\phi}$.
It is convenient to write
$P^{+} = e^{-i (2\pi/N) \hat n}$,
where $n$ can be thought of as the number variable conjugate
to the phase.  Note that $\psi^\dagger$ indeed acts as
a raising operator on this number as suggested by the notation.


\end{thebibliography}
\end{document}